\documentclass[conference]{IEEEtran}
\IEEEoverridecommandlockouts
\usepackage{cite}
\usepackage{amsmath,amssymb,amsfonts}
\usepackage{algorithmic}
\usepackage{graphicx}
\usepackage{balance}
\usepackage{textcomp}
\usepackage{xcolor}
\usepackage{url}
\def\BibTeX{{\rm B\kern-.05em{\sc i\kern-.025em b}\kern-.08em
		T\kern-.1667em\lower.7ex\hbox{E}\kern-.125emX}}
\begin{document}
	
	\title{What's in a GitHub Repository? - A Software Documentation Perspective\\}

	
\author{\IEEEauthorblockN{Akhila Sri Manasa Venigalla, Sridhar Chimalakonda}
\IEEEauthorblockA{\textit{Research in Intelligent Software \& Human Analytics (RISHA) Lab}\\
\textit{Department of Computer Science and Engineering} \\
\textit{Indian Institute of Technology Tirupati}\\
Tirupati, India\\
\{cs19d504, ch\}@iittp.ac.in }
}
	\maketitle
	
	\begin{abstract}
	Developers use and contribute to  repositories on GitHub. Documentation present in the repositories serves as an important source by helping developers to understand, maintain and contribute to the project. Currently, documentation in a repository is diversified, among various files, with most of it present in ReadMe files. However, other software artifacts in the repository, such as \textit{issue reports} and \textit{ pull requests} could also contribute to documentation, without documentation being explicitly specified.
		Hence, in this paper, we propose a \textit{taxonomy of documentation sources} by analyzing different software artifacts, developer interviews and card-sorting approach. We inspected multiple artifacts of 950 public GitHub repositories, written in four different programming languages, C++, C\#, Python and Java, and analyzed the type and amount of documentation that could be extracted from these artifacts. To this end, we observe that, about 25.93\% of information extracted from all sources proposed in the taxonomy contains \textit{error-related} documentation, and that \textit{pull requests} contribute to around 18.21\% of extracted information.\\
	\end{abstract}
	
	\begin{IEEEkeywords}
		Software Artifacts, GitHub, Documentation, Issue Reports, Pull Requests, Commit Messages
	\end{IEEEkeywords}
	
	\section{Introduction}
	GitHub is one of the leading platforms of open source software projects \cite{gousios2014lean}. 
	GitHub facilitates several developers to collaborate and contribute to projects by performing several actions such as updating projects using commits, viewing changes made to projects through commit history \cite{tsay2014influence}, logging issues or defects through issue reports and contributing to other projects through pull requests and so on \cite{jiang2017and, jiang2017understanding}. Open source projects originating from well known organizations such as \textit{vscode} from Microsoft, \textit{react-native} from Facebook and \textit{tensorflow} from Google, have more than 1K\footnote{\url{https://github.com/microsoft/vscode}}, 2K\footnote{\url{https://github.com/facebook/react-native}} and 2.8K\footnote{\url{https://github.com/tensorflow/tensorflow}} contributors respectively (as of January 2021). Developers also contribute to and reuse several other publicly available repositories on GitHub \cite{zagalsky2015emergence}. Documentation in the repositories facilitates developers to understand about a project and consequently helps them in deciding on projects they wish to contribute \cite{borges2018s}. Repositories contain \textit{Readme} files that provide information on the purpose, requirements, usage instructions and various other information about the repositories \cite{prana2019categorizing, perez2016ten}. Several other files of the repository such as \textit{License} files, \textit{UML} files and so on also present different types of repository information such as license permissions and design decisions \cite{hebig2016quest, vendome2015license}.
	Though many developers are interested contribute to GitHub repositories, they face multiple hurdles during this process, resulting in reduced motivation towards contributing to repositories \cite{mendez2018open, steinmacher2016overcoming}. The existing insufficient and scattered documentation on these repositories makes it difficult for developers to understand about the repositories, consequently reducing the advantages of huge number of contributions \cite{aghajani2019software}. Developers tend to visit other artifacts in the repositories to better understand about the repository, which is effort-intensive \cite{fronchetti2019attracts, robles2006beyond}. We believe that contribution efforts from wide range of developers could be well leveraged if documentation is improved and consolidated.
	
	Researchers have explored multiple dimensions of understanding, usage and ways to improve software documentation, as it plays a significant role in performing various tasks. These tasks include software development, testing, integration, maintenance and so on \cite{garousi2015usage, mahmood2011industrial}, but, the documentation is insufficient in majority of the projects \cite{aghajani2020software, aghajani2019software}. 
	For example, documentation helps in improving software development and maintenance by providing necessary information to users in different roles such as system integrators, quality analysts, developers and so on \cite{kipyegen2013importance}. Good documentation helps in re-engineering existing software during maintenance and migration \cite{de2005study, aghajani2020software}. It has also been observed that developers regard documentation as an important aspect, even in projects with faster release cycles such as agile projects \cite{stettina2011necessary}. 
	
	Considering the wide usage of GitHub and activities performed by developers on GitHub, documentation of software repositories that are hosted on GitHub also plays a major role in various stages of the project such as development, deployment, maintenance and so on. It has been observed that projects having better popularity tend to have better documentation \cite{cosentino2017systematic}. Developers of popular projects tend to improve documentation by regular updates, to provide better insights to users \cite{aggarwal2014co}. Also, documentation is one of the factors considered by developers before contributing to repositories on GitHub \cite{borges2018s}. Currently, documentation is spread across multiple files such as \textit{source code}, \textit{design diagrams}, \textit{readme files}, \textit{license files} and so on \cite{hebig2016quest, prana2019categorizing, ma2018automatic}. We hypothesize that other artifacts could also provide valuable information about a repository, that could be considered as documentation, without documentation being explicitly present. Researchers have also observed that developers spend considerable effort on multiple artifacts, other than source code, in end-user oriented projects, supporting our idea of considering documentation from multiple sources \cite{robles2006beyond, fronchetti2019attracts}. Currently, sources of documentation are diversified in a repository. The documentation present in these sources is also unclear. \textit{Unifying the information from multiple sources could help in enhancement of documentation, and eventually reduce effort for developers, thus motivating them to contribute to GitHub repositories.} Though there are studies to identify different types of documentation present in software repositories \cite{prana2019categorizing, hebig2016quest, chimalakonda2020software}, we are not aware of any work that integrates documentation from multiple artifacts, thus, motivating our work. Hence, in this paper, we aim to identify and gather different artifacts of software repositories, that could contain information relevant to documentation. The main contributions of this paper are:
	\begin{itemize}
		\item A taxonomy of documentation sources in GitHub Repositories based on card-sorting approach and developer-based interviews with 20 developers.
		
		\item An empirical analysis on 950 GitHub repositories, of four programming languages, C++, C\#, Python and Java, to understand the types of documentation that could be extracted from multiple software artifacts.
		
		\item Results of the empirical study, along with the percentage of available information that could contribute to documentation, in each of the software artifacts that are identified as potential sources of documentation. The results of the study and dataset used for the study can be accessed here\footnote{Results \& Dataset - \url{https://osf.io/dfx9r/?view_only=0954174b44054893a3dfbe6cc7dd5db5}}.
		\end{itemize}
	\vspace{1mm}		     \fbox{\begin{minipage}{25em}
	\textit{\textbf{Documentation Type}} - We define type of documentation based on content present in the documentation. For example, documentation that refers to reporting or fixing an error is considered as \textit{Error-related} documentation type.
	\end{minipage}
	}
	\vspace{1mm}
		\newline
	     \newline
		     \fbox{\begin{minipage}{25em}
	\textit{\textbf{Documentation Source} - } We define documentation source as a software artifact of the project, capable of providing potentially useful information related to software documentation. 
	\end{minipage}
	}


	\section{Research Methodology}
	In this paper, we aim to analyze contents in GitHub repositories from a software documentation perspective, specifically focusing on various software artifacts that could serve as sources of documentation. 
	We followed an approach that comprises of the following six phases.
	\begin{itemize}
	
	\begin{figure*}
	    \centering
	    \includegraphics[width = \linewidth]{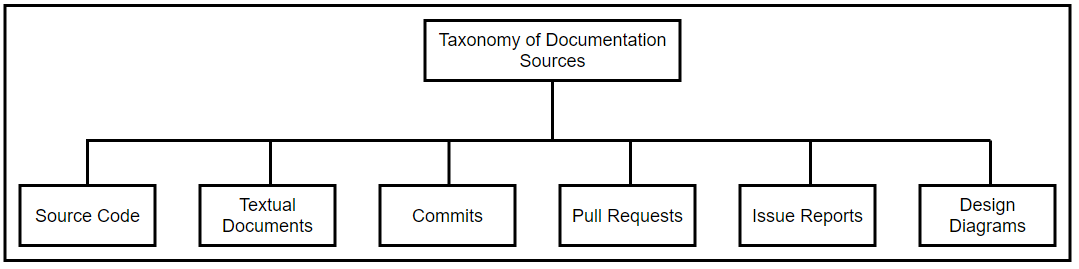}
	    \caption{Proposed Taxonomy of Documentation Sources}
	    \label{fig:taxonomy}
	\end{figure*}
	
		\begin{figure*}
	    \centering
	    \includegraphics[width = \linewidth]{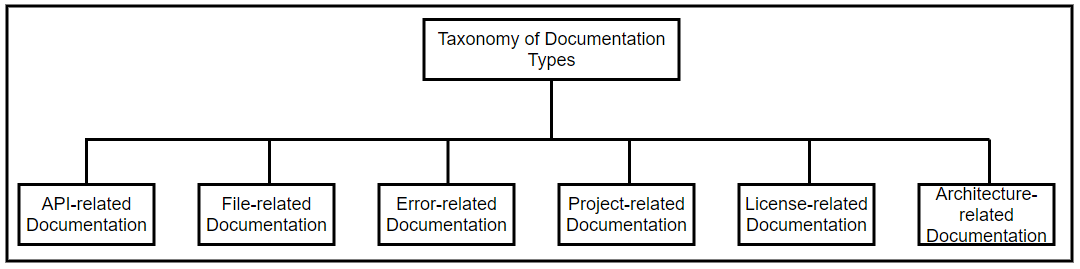}
	    \caption{Proposed Taxonomy of Documentation Types}
	    \label{fig:typestax}
	\end{figure*}
	    \item \textit{Research Questions Definition} - Four research questions aimed to identify possible sources of software documentation have been defined.
		
		\item \textit{Card Sorting Approach and Developer Interviews} - For manual analysis, 60 of the 950 GitHub repositories, which include 15 top trending repositories from C++, C\#, Java and Python programming languages, were selected analyzed by three individuals using card sorting approach to narrow down onto potential categories of documentation and sources that could be leveraged in answering the research questions. Twenty professional software developers were interviewed to understand different types of documentation and sources as perceived by developers.
		
		\item \textit{Data Extraction} - 950 public GitHub repositories, which include 233, 241, 253 and 223 repositories from C++, C\#, Java and Python programming languages respectively, were scraped and processed for further analysis. 
		
		\item \textit{Topic Modelling and Manual Inspection} - LDA topic modelling techniques were applied to 20\% of the scraped data to identify topics in the data. The topics identified are manually walked through to label the categories.
		
		\item \textit{Automated Analysis} - Models to analyze extracted data against research questions are developed.
		
		\item \textit{Result Comprehension} - Research questions are answered as a result of automated analysis.
	\end{itemize}
	
	\subsection{Research Questions}
	The main goal of this study is to propose a taxonomy of documentation sources in GitHub, by identifying different potential sources of software documentation in a GitHub repository, including the sources in which documentation is not explicitly specified. We conduct this study by answering the following research questions:
	
	\begin{itemize}
	    \item \textbf{RQ1: What are the types of documentation present in GitHub repositories ?} 
	    
	    To answer this question, we employed card-sorting approach and developer interviews with 20 developers to understand different documentation types present in GitHub repositories, as perceived by GitHub users.
	    
	    \textbf{Summary:} We observe that majority of the GitHub repositories comprise of six documentation types, of which at least three of them are not explicitly mentioned as documentation in the repositories.
	    
	    \item \textbf{RQ2: What are the sources of documentation in GitHub repositories? }
	    
	    We answer this question by identifying various sources that are perceived to contribute to different documentation types through developer interviews with 20 developers and card-sorting approach with 3 participants. 
	    
	    \textbf{Summary:} We observe that majority of the developers use six artifacts of GitHub Repositories, to understand about a repository, and further contribute to it. This also included \textit{issues} and \textit{pull requests}, which do not explicitly convey documentation related information.
	    
	   	\item \textbf{RQ3: What is the distribution of different documentation types in each of the identified sources?}
	    
	    To answer this question, we categorize the information extracted from each source of 950 GitHub repositories into one of the documentation types identified, and calculate the percentage of each type in the sources.
	    
	    \textbf{Summary:} Of the six documentation types identified, \textit{error-related} documentation is present to a larger extent in majority of the documentation sources.
	    
	    \item \textbf{RQ4: What is the contribution of identified sources to different documentation types?}
	    
	    To answer this question, we calculate the percentage contribution of a source by finding the ratio of amount of information present in a source with respect to the documentation type to the total amount of information of the documentation type, present across different sources.
	    
	    \textbf{Summary:} Following a well-established notion, majority of the documentation was observed to be present in \textit{source-code} comments and \textit{textual documents}. However, \textit{commit logs} and \textit{issues} were also observed to contribute to documentation, with a minor variation in the percentage, in comparison to the \textit{textual} documentation.

	    

 	\end{itemize}
	
\subsection{Card Sorting Approach and Developer Interviews}
We observe that card-sorting \cite{spencer2004card} approach has been used widely towards arriving at various taxonomies and classifications in the software engineering literature, such as classifying requirements change \cite{nurmuliani2004using} and research ideas\cite{lo2015practitioners}, which motivated us to employ card-sorting as an initial step towards arriving at different types of documentation and their sources.

We have downloaded 60 of the 950 GitHub repositories, such that they include 15 top trending repositories, from each of the four programming languages being considered, for manual analysis. Three individuals, comprising two researchers and one under-graduate student were assigned the task of identifying different types and sources of documentation present in the repositories. Initially, all the three individuals have explored \textit{readme files} of the repositories, considering them to be a basic and easily available source of documentation. We employed open card sorting approach, without pre-defining possible groups of documentation. In the first step, each of the three individuals came up with different number of groups (4, 6 and 10). 

On further iterative discussions on labels of each of the groups, we observed similarities between the information content present among the 10 groups identified by the third individual, resulting in a set of 5 groups (T1 to T4 and T6 documentation types presented in Table \ref{tab:typedefs}). We further compared and discussed about the data content among other groups, identified by each individual and finally arrived at a common decision, with 6 possible types of documentation, accompanied by labels to these types, as mentioned in Table \ref{tab:typedefs}. As a next step, all the three individuals navigated through various artifacts within the repositories, and attempted to identify documentation in those artifacts. On identification of documentation, each artifact was labelled with the corresponding documentation type (identified in the previous step) it would be representing. While one individual narrowed down to five artifacts, the other two individuals pointed out six possible artifacts containing documentation. Each of the three individuals have explained the reasons for selecting respective number of artifacts, based on which, we finally decided to consider six different artifacts (mentioned in Fig. \ref{fig:taxonomy}) as part of the proposed taxonomy.

We later organised developer interviews with 20 professional developers to validate the types and sources of documentation obtained as a result of the card sorting approach. We approached 20 developers working in different organizations, with work experience ranging from 2 to 30 years. All the 20 developers interviewed belonged to organizations with more than 1000 employees and with a team size of at least 10. We selected developers with this profile to ensure that they are well-equipped with the collaborative project development and management. All the 18 developers with less than 10 years of experience mentioned that they actively contribute to open-source projects. 
We asked them open-ended questions to understand their perception about types of documentation and the sources of documentation present on GitHub.  12 of the developers, with work experience ranging between 3 to 10 years explained that they find six documentation types (presented in Table \ref{tab:typedefs}) in the projects they interact with, on GitHub.  2 developers with work experience of 29 and 30 years pointed out that they find seven types of documentation in the projects, which included \textit{migration guidelines}, apart from the six types identified in Table \ref{tab:typedefs}. \textit{Migration guidelines} could however be considered as \textit{project-related} documentation, at a higher level of abstraction, as they specify details necessary in migrating the project. The rest of the developers with less than three years of experience have expressed that they have observed only three types of documentation (T1, T2 and T5 in Table \ref{tab:typedefs}). This varied insights from developers could also be due to varied abstractions of projects they are exposed to, in their respective organizations. We then discussed with them the documentation types identified previously, through card-sorting approach. 18 of the 20 developers have later agreed to the identified six documentation types, while the other two developers have suggested inclusion of one more category, which corresponds to \textit{migration guidelines} of the project. 

Considering the majority responses, we decided to proceed further, with six documentation types. We further queried the developers to understand different artifacts on GitHub that they use to understand about a repository, which could be helpful, if collated as a single documentation file. All the 20 developers have pointed out six sources of documentation, as shown in Figure \ref{fig:taxonomy},  in GitHub repositories, and also mentioned that five (\textit{Source code}, \textit{textual documents}, \textit{commits}, \textit{pull requests} and \textit{issue reports}) of these six documentation sources, are frequently referred by them.

\begin{table}[]
\caption{Observed Documentation Types and corresponding definitions }
\vspace{-2mm}
    \centering
    \begin{tabular}{|l|l|l|}
	          \hline
	          \textbf{S.}&
	          \textbf{Types of} & \textbf{Definition}\\
	          \textbf{no.}& \textbf{Documentation} & \\
	          \hline
	          T1 & API-related & Documentation capable of providing information  \\
	          &documentation& about APIs used in the project\\
	          \hline
	          T2 & File-related & Documentation capable of providing \\
	          &Documentation & file-level information such as updates made to \\
	          &&the files and dependencies of files\\
	          \hline
	          T3 & Project-related  & Documentation capable of providing \\
	          &Documentation & project-level information such as installation \\
	          & & instructions, branches in projects, enhancements \\
	          & & to the projects and so on\\
	          \hline
	          T4 & License-related & Documentation capable of providing information \\
	          &Documentation & about licenses in the project
	          \\
	          \hline
	          T5 & Error/Bug-related  & Documentation capable of providing information \\
	          &documentation& about errors or bugs encountered \\
	          & & in multiple files of project\\
	          \hline
	          T6 & Architecture- &  Documentation capable of providing information \\
	          &related  &about architecture of the project, such as \\
	          &Documentation&module interactions.\\
	          
	          \hline
	    \end{tabular}
    
    \label{tab:typedefs}
     \vspace{-2mm}
\end{table}
	
	Based on the card sorting approach and developer interviews, we propose six potential documentation sources, that includes 3 software artifacts which do not have documentation explicitly specified, as shown in Fig. \ref{fig:taxonomy}. Thus, we observe six types of documentation and six sources of documentation, as mentioned in Table \ref{tab:typedefs} and Fig. \ref{fig:taxonomy} respectively.
		    \newline
		     \newline
		     \fbox{\begin{minipage}{25em}
		    \textbf{RQ1:} We observed six different documentation types in GitHub Repositories. \textit{API-related}, \textit{file-related}, \textit{project-related}, \textit{license-related}, \textit{error/bug-related} and \textit{architecture-related} documentations are the six different documentation types identified.
		    
		    \textbf{Insights-} Developers perceived \textit{file-related} and \textit{project-related} documentation types to be more useful in contributing to a project, among the six types.
		    
		     \end{minipage}}
		     \vspace{1mm}
		     \newline
		     \newline
		    \fbox{\begin{minipage}{25em}
		    \textbf{RQ2:} We observed six different sources containing potential information related to documentation in GitHub repositories.\textit{ Source code, textual documents, commits, pull requests, issue reports and design diagrams} are the six artifacts, observed to be potential sources of documentation in GitHub repositories.
		    
		    \textbf{Insights-} Apart from the widely accepted documentation sources - \textit{source code comments},\textit{design diagrams} and \textit{textual documents}, our interviews with developers revealed that \textit{pull requests}, \textit{commits} and \textit{issues} can also contribute to documentation. 
		     \end{minipage}}
		     
	    	    

	\subsection{Data Extraction}
	We identified top-starred 300 GitHub repositories, written in each of the four different programming languages, C++, Python, C\# and Java. The choice of languages was based on random numbers generated from 1 to 10, and the languages corresponding to these random numbers among the top 10 active languages on GitHub\footnote{https://githut.info/}, as of 01 
	January 2021. We further filtered out forked repositories and other repositories with zero pull requests, to exclude tutorial-based repositories. This resulted in 233, 241, 253 and 223 repositories corresponding to C++, C\#, Java and Python programming languages respectively, leading to a total of 950 repositories. We analyzed each of the repositories to extract information from multiple documentation sources. We have manually walked through top 15 repositories in each of the programming languages to identify file formats for different documentation types. Based on the findings from developer interviews and card-sorting approach, we classified all files in the repositories into 5 categories - Textual Documentation, Images, Design Diagrams, Source Code and Others. The rules employed for this categorization are presented in the Table \ref{tab:fileclass}.
	\begin{table}[]
	    \caption{File Classification rules}
	    \centering
	    \begin{tabular}{|c|c|}
	    \hline
	   \textbf{Category }  &  \textbf{File Format} \\
	   \hline
	   Textual & `.txt' , `.md' , `readme' , `license'\\
	   Images & `.png', `.jpg', `.jpeg'\\
	   Design Diagrams  & `.xmi',`.uml'\\
	   Source Code & `.cpp', `.cs', `.py', `.java'\\
	   Others & Any other extensions\\
	   \hline
	   \end{tabular}
	    \label{tab:fileclass}
	\end{table}
	
	Files with extensions - {`.txt', `.md', } are categorized to \textit{textual} documentation.
	All files with extensions - {`.xmi', `.uml'} are categorized to UML diagrams.
	Files with {`.jpg',`.jpeg',`.png'} extensions are primarily categorized as Images, which could be categorized as design diagrams through further analysis. Files containing extension of programming languages considered, i.e., .cpp, .py, .java and .cs are classified as source code files. Though the repositories were extracted based on the programming language in which they are written, a majority of the repositories also contain files written in other programming languages. 
	
	Considering the vast number of programming languages supported by GitHub, it is difficult to analyze files of other programming languages, as programming constructs differ from one language to another. Hence, files with formats not belonging to the above four classes are categorized as \textit{Others}. Comments from files in the \textit{source code} category have been extracted. This exploration of repository files has extracted information that could contribute to documentation, from two documentation sources mentioned in the taxonomy, namely - \textit{textual documents} and \textit{source code files}. Also, \textit{design diagrams} for each of the repositories have been identified and information from these diagrams could be extracted using image processing techniques. In view of the technical effort and difficulty involved in choosing and applying appropriate image processing techniques, we consider data extraction from design diagrams to be out of the scope of this study.
	
	For each of the repositories, other software artifacts, i.e., \textit{pull requests, issues} and \textit{commits}, that were identified as sources of documentation in the proposed taxonomy, not older than three years from January 01, 2021,  were extracted using GitHub API. However, blank data was returned for 46 python repositories, leaving us with pull request data of only 177 of the 223 Python repositories considered. This blank data could be due to the large number of pull requests in these repositories, which could not be extracted due to rate limits of GitHub API, or those repositories with pull requests older than three years, from 01 January, 2021. 
	
We have identified fields that are capable of containing documentation related information through manual inspection of extracted artifacts in 50 of the 950 repositories, selected randomly. These fields are presented in Table \ref{tab:fields}, and the data of these fields is stored as text documents. This accomplishes the task of extracting information useful for documentation from three sources of documentation, mentioned in the proposed taxonomy, namely  - \textit{Pull Requests, Issues} and \textit{Commits}.
\begin{table}[]
    \caption{Documentation Sources and fields considered}
    \centering
    \begin{tabular}{|c|c|}
    \hline
    \textbf{Documentation Source} & \textbf{Fields Identified}\\
    \hline
    Issues &  title , body, comments\\
    Pull Requests & title, body, comments\\
    Commits & message, comments\\
    \hline
    \end{tabular}
   
    \label{tab:fields}
\end{table}
The fields that contain textual data and contribute to documentation types are identified (as presented in Table \ref{tab:fields}) and corresponding text is stored for further analysis.


	\subsection{Topic Modelling and Manual Inspection}
	We observed that software artifacts that are identified as sources of documentation contain information that could support different documentation types. Hence, we analyze this information and categorize into different documentation types. Topic modelling technique can identify keywords, that belong to different number of topics specified, in a given document. This feature of topic modelling fits well with our requirement of categorizing text into different documentation types. We performed Latent Dirichlet Allocation (LDA) topic modelling technique to categorize the information. Though we have consolidated different types of documentation, all sources of documentation need not contribute to all types of documentation. Hence, we primarily tried to understand the optimal number of topics into which information from each of the software artifacts could be classified. Towards this, LDA models with topics in the range of 2 to 20 have been generated\footnote{LDA has been applied on data of each artifact individually, for each repository}, for textual data from each of the software artifacts for 50 randomly selected repositories in each programming language. Coherence scores for these models have been calculated and the number of topics of model with highest coherence score are identified as optimal number of topics for the corresponding data source. It is to be noted that data of different repositories or different artifacts was not combined during this process, to ensure that the topic number is not influenced by the data of other repositories. We observed that the number of optimal topics for issue data varied between 4 and 5, with 5 having higher frequency than 4. The optimal topics for commit data and pull request data were also observed to be varying between 4 and 5, but, with 4 being more frequently repeated, than 5. Hence, the optimal number of topics for commit data, issue data and pull request data are 4, 5 and 4 respectively. LDA topic modelling was applied on the same data\footnote{Same data used to identify optimal number of topics} with optimal number of topics and the top 10 keywords for each of the topics were obtained. 
	
	A manual inspection of the keywords obtained for each artifact has been performed repository wise, for the 200 repositories (50 from each programming language) being considered. We have manually walked through keywords in each topic, compared them to source code of the repository and other files in the repository to identify the object of reference, and labelled the topics accordingly. During the manual inspection of keywords, we did not find any keywords that explicitly correspond to architecture-related documentation. Considering the design diagrams to contain more relevant architecture related documentation, we assumed such documentation to be absent in the documentation sources being considered for this study. Based on the identified keywords, it could also be observed that though these sources might contain architecture-related documentation, its presence is almost negligible, thus, resulting in a set of 5 documentation types being present in the considered documentation sources. 
	Though the optimal number of keywords for \textit{pull request} data and \textit{commit} data was observed to be 4 through LDA approach, we observed the data to contain keywords from the 5\textsuperscript{th} category as well during our manual analysis. This indicates that though the information present in \textit{pull requests} and \textit{commit} data can be classified into 4 categories, the set of these four categories, differ across repositories. For example, some repositories might comprise of \textit{API-related}, \textit{File-related}, \textit{Project-related} and \textit{Error-related} documentation, while some repositories might contain \textit{License-related, File-related, Project-related} and \textit{Error-related} documentation in the \textit{pull request} and \textit{commit} data. 
	
	While labelling, in some cases, more than one topics were observed to have same labels, which indicate that different repositories have varied distributions of information in view of documentation types. Keywords of similar topics identified for all the artifact data are compared and 10 most frequently repeated keywords for each documentation type are identified. 	
	
	\subsection{Automated Analysis}
	
	A rule based classification model is built by considering the 10 most frequent keywords of each topic and its corresponding label.  
	The topics obtained as a result of topic modelling for each software artifact are labelled using the rule based classifier.
	The rule based classifier includes rules to label topics based on similarity score of the topic with respect to keyword sets of each of the 5 identified categories(\textit{error-related}, \textit{file-related}, \textit{project-related}, \textit{license-related} and \textit{API-related}). Topics that have almost equal similarity scores (difference less than 0.05) for all categories are labelled into \textit{others} category. 
	Thus, artifact data having almost equal possibility to belong to more than one categories are classified into \textit{others} category.
	
	The percentage of each documentation type is calculated based on the topic frequency in the artifact. These two features - identifying documentation types and calculating percentages, are integrated into a result generator script written in python programming language. This result generator takes as input list of information extracted from multiple software artifacts of all 950 repositories and automatically generates percentage of different documentation types present in each of the software artifacts for all the 950 repositories. Thus, the percentages generated imply frequency of related-keywords of each documentation type, in each of the artifacts.
	

    \section{Results}
\begin{figure}[h!]
        \centering
        \includegraphics[width = \linewidth]{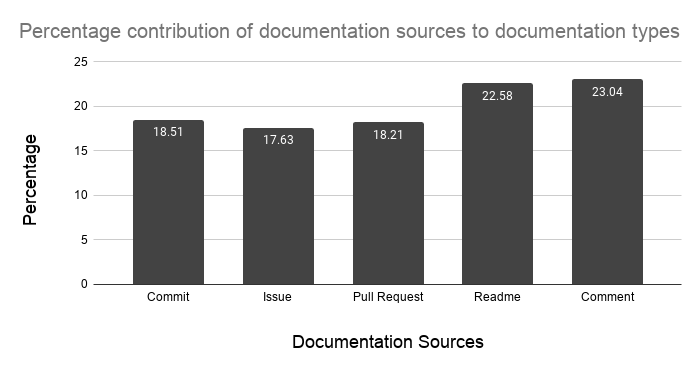}
        \caption{Percentage of contribution of documentation sources to all documentation types for all 950 repositories.}
        \label{fig:sources}
        \vspace{-4mm}
    \end{figure}

    The distribution of documentation types obtained as a result of automated analysis of each artifact data is presented in Fig. \ref{fig:types}. The contribution of sources to all documentation types is presented in Fig. \ref{fig:sources}.
    Also the results of artifact contributions to specific documentation types and distribution of documents among specific sources across multiple programming languages are presented in the form of plots.
	
	\subsection{Distribution of Documentation Types in Documentation Sources}
	\begin{itemize}
	    \item \textbf{Source Code Comments -} The percentage distribution of all documentation types in \textit{Source code comments} is presented in Fig. \ref{fig:comments}. It has been observed that majority of the information present in source code comments contribute to license based documentation, in repositories of all programming languages considered. Also, we observed that information present in source code comments contribute the least to \textit{error-related} documentation, in repositories of C\# and Java programming languages, while that in repositories of C++ and Python programming languages contribute the least to \textit{API-related} documentation.  
		    	    
	\begin{figure}[h!]
	    \centering
	    \includegraphics[width = \linewidth]{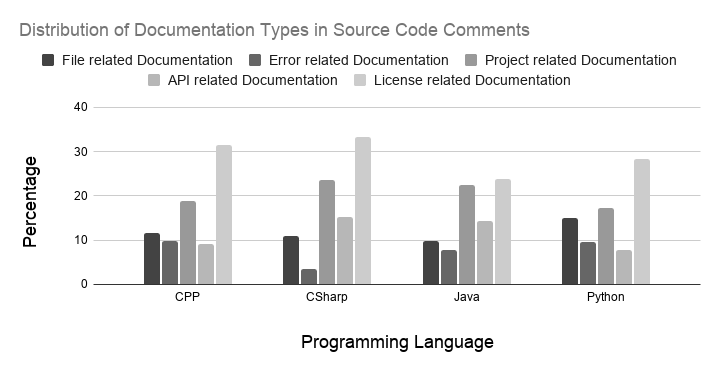}
	    \caption{Distribution of document types in source code comments in repositories of the four programming languages considered.}
	    \label{fig:comments}
	    \vspace{-4mm}
	\end{figure}
	\begin{figure}[h!]
	    \centering
	    \includegraphics[width = \linewidth]{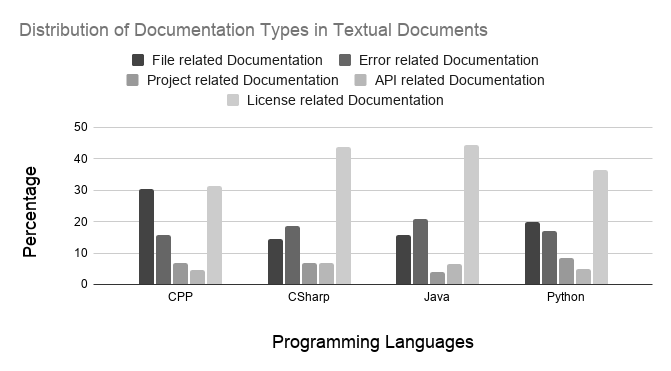}
	    \caption{Distribution of document types in textual documents in repositories of the four programming languages considered.}
	    \label{fig:texts}
	    \vspace{-4mm}
	\end{figure}
	\begin{figure}[h!]
	    \centering
	    \includegraphics[width = \linewidth]{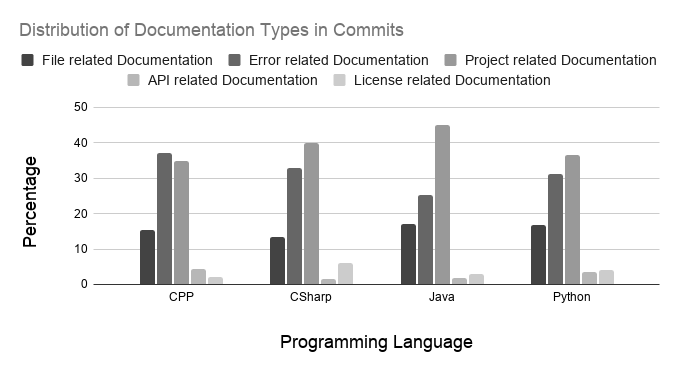}
	    \caption{Distribution of document types in commits in repositories of the four programming languages considered.}
	    \label{fig:commits}
	    \vspace{-4mm}
	\end{figure}
	\begin{figure}[h!]
	    \centering
	    \includegraphics[width = \linewidth]{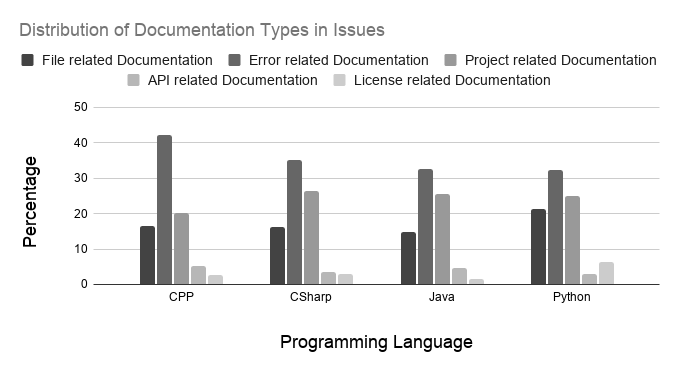}
	    \caption{Distribution of document types in issues in repositories of the four programming languages considered.}
	    \label{fig:issues}
	    \vspace{-4mm}
	\end{figure}
	\begin{figure}[h!]
	    \centering
	    \includegraphics[width = \linewidth]{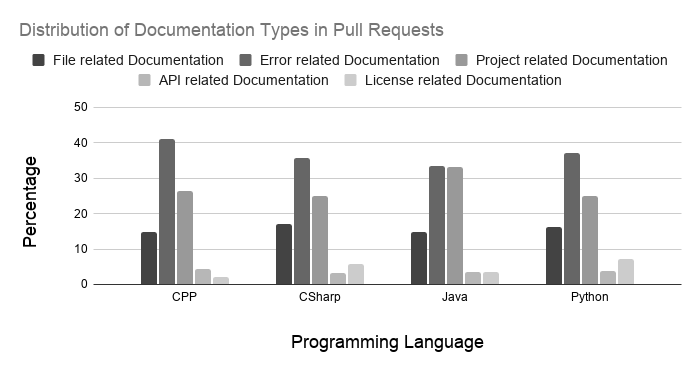}
	    \caption{Distribution of document types in pull requests in repositories of the four programming languages considered.}
	    \label{fig:prs}
	    \vspace{-4mm}
	\end{figure}
	\begin{figure}[h!]
	    \centering
	    \includegraphics[width = \linewidth]{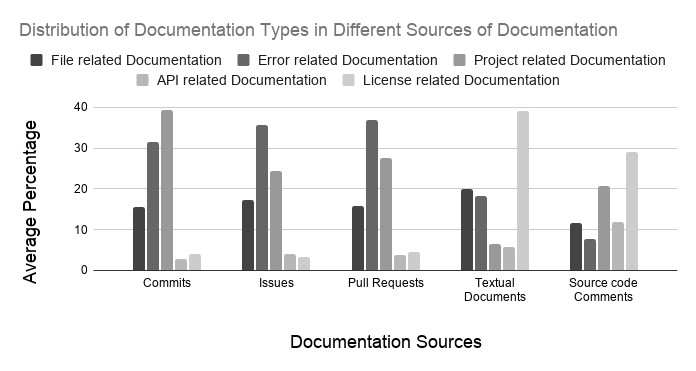}
	    \caption{Average distribution of document types in all sources, across all 950 repositories.}
	    \label{fig:all}
	    \vspace{-4mm}
	\end{figure}

	    \item \textbf{Textual Documents-} The plot shown in Fig. \ref{fig:texts} indicates that most of the information in \textit{textual documents} of the repositories contribute to \textit{license-related} documentation, followed by \textit{file-related} documentation in repositories of C++ and Python programming languages.
	    
	    \item \textbf{Commits-} The plot displayed in Fig. \ref{fig:commits} indicates that most of the information present in textual sentences of commit logs contribute to \textit{project-related} documentation in majority of the repositories, followed by \textit{error-related} documentation. For repositories in C++ programming language, majority of the information in commit logs contributes to \textit{error-related} documentation, followed by \textit{project-related} documentation.
	    
	    \item \textbf{Issues-} The plot shown in Fig. \ref{fig:issues} indicates that percentage distribution of \textit{error-related} documentation and \textit{project-related} documentation are prominent in textual sentence information of issues, with \textit{error-related} documentation having highest percentage distribution across repositories in all 4 programming languages.
	    
	    \item \textbf{Pull Requests-} Fig. \ref{fig:prs} shows that most of the information in textual sentences of pull requests could potentially contribute to \textit{error-related} documentation, with all repositories having an average of more than 30\% information that contributes to \textit{error-related} documentation. 
	    \end{itemize}
		    A consolidated distribution of documentation types across all identified sources of documentation is presented in Fig. \ref{fig:all}, which indicates that minimal amount of information in \textit{commits}, \textit{issues} and \textit{pull requests}, and majority of information in \textit{textual} \textit{documents} and\textit{ source code comments} contributes to \textit{license based} documentation.

\begin{figure}[h!]
        \centering
        \includegraphics[width = \linewidth]{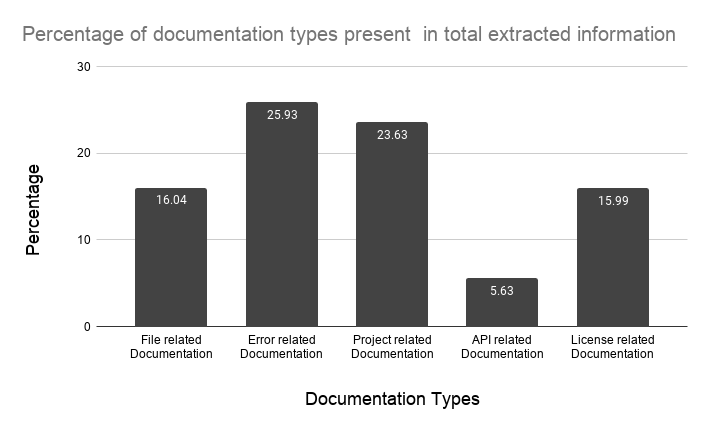}
        \caption{Percentage of documentation types present in information extracted from all sources of documentation for all 950 repositories}
        \label{fig:types}
        \vspace{-4mm}
    \end{figure}

		     Fig. \ref{fig:types} indicates that \textit{error-related} documentation largely exists in the information extracted.
		     We further observed that about 12.75\% of the total information did not contribute to any of the identified documentation types.
		     \newline
		     \newline
		     \fbox{\begin{minipage}{25em}
		    \textbf{RQ3:} Majority of the total extracted documentation from multiple sources, consists of \textit{error-related} documentation (25.9\%), followed by \textit{project-related} documentation (23.6\%). \textit{File-related} documentation and \textit{License-related} documentation account to 16.04\% and 15.99\% of the total extracted documentation respectively. 5.63\% of the total extracted documentation consists of \textit{API-related} documentation. 
		    
		    \textbf{Insights-} Tagging information across documentation sources, based on corresponding documentation types could help developers in better comprehension of the project. 
		     \end{minipage}}
	\begin{figure}[h!]
	    \centering
	    \includegraphics[width = \linewidth]{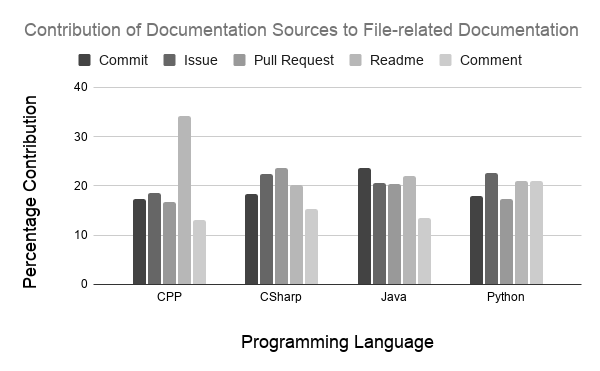}
	    \caption{Contribution of documentation sources to \textit{file-related} documentation in repositories of the four programming languages considered}
	    \label{fig:file}
	    \vspace{-4mm}
	\end{figure}
	\begin{figure}[h!]
	    \centering
	    \includegraphics[width = \linewidth]{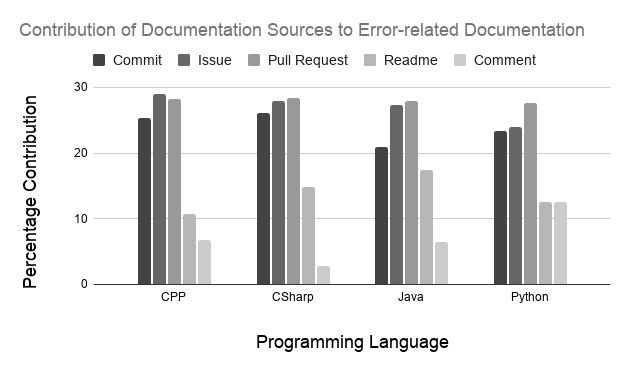}
	    \caption{Contribution of documentation sources to error/bug-related documentation in repositories of the four programming languages considered}
	   \label{fig:error}
	    \vspace{-4mm}
	\end{figure}
	\begin{figure}[h!]
	    \centering
	    \includegraphics[width = \linewidth]{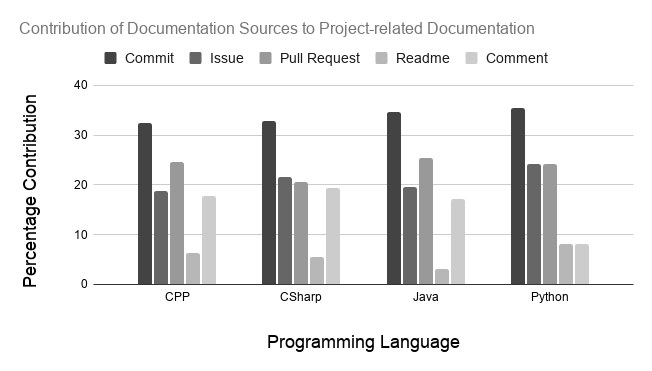}
	    \caption{Contribution of documentation sources to \textit{project-related} documentation in repositories of the four programming languages considered}
	    \label{fig:project}
	    \vspace{-4mm}
	\end{figure}
	\begin{figure}[h!]
	    \centering
	    \includegraphics[width = \linewidth]{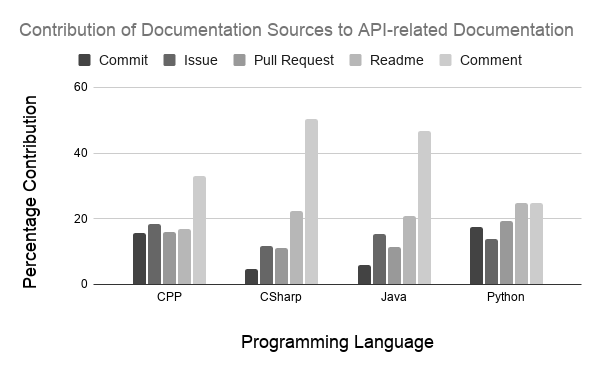}
	    \caption{Contribution of documentation sources to \textit{API-related} documentation in repositories of the four programming languages considered}
	    \label{fig:api}
	    \vspace{-4mm}
	\end{figure}
    \begin{figure}[h!]
	    \centering
	    \includegraphics[width = \linewidth]{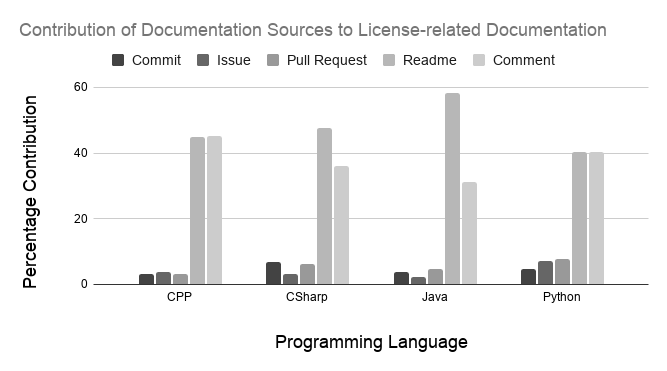}
	    \caption{Contribution of documentation sources to \textit{license-related} documentation in repositories of the four programming languages considered}
	    \label{fig:lic}
	    \vspace{-4mm}
    \end{figure}
	    \begin{figure}[h!]
	    \centering
	    \includegraphics[width = \linewidth]{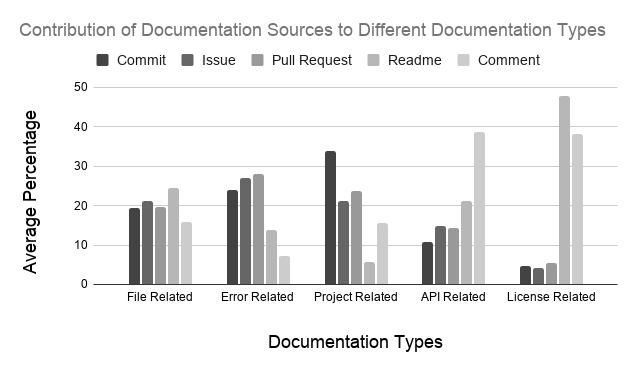}
	    \caption{Average contribution of documentation sources to all documentation types, across all 950 repositories}
	    \label{fig:types1}
	    \vspace{-4mm}
    \end{figure}

\subsection{ Contribution of Documentation Sources to Documentation types}
	    \begin{itemize}
	        \item \textbf{File-related Documentation-} The plot in Fig. \ref{fig:file} represents contribution of each documentation source to \textit{file-related} documentation, which varied across repositories of the 4 programming languages. We observe that \textit{pull requests} contribute to most of the \textit{file-related} documentation in repositories of C\#, while \textit{textual documents} contribute the most in repositories of C++ programming language. We also observed that source-code comments contribute the least to \textit{file-related} documentation in repositories of C++, C\# and Java programming languages.
	        
	        \item \textbf{Error-related Documentation-} Fig. \ref{fig:error} shows that maximum contribution of \textit{error-related} documentation is from \textit{pull requests} and \textit{issues}. It can also be observed that \textit{textual} documents and \textit{source-code comments} contain the least amount of potential information with respect to \textit{error-related} documentation.
	        \item \textbf{Project-related Documentation-} The plot in Fig. \ref{fig:project} indicates that most of the \textit{project-related} documentation is contributed by information in \textit{commit logs}, when compared to other artifacts being considered.

	        \item \textbf{API-related Documentation-} We deduce that source-code comments in repositories contain more information about \textit{API-related} documentation, succeeded by textual documents in the repositories from the plot in Fig. \ref{fig:api}.
	        
	        \item \textbf{License-related Documentation-} It could be observed from the plot shown in Fig. \ref{fig:lic}, that textual documents in majority of the repositories in all four programming languages contribute to \textit{license-related} documentation, followed by source code comments in repositories. 
	        
	    \end{itemize}

	    The consolidated contribution of documentation sources towards all documentation types is presented in Fig. \ref{fig:types1}, which indicates that most of the potential information for \textit{file-related} documentation 
	    is obtained by data in \textit{textual} documents.
	    
	    Fig. \ref{fig:sources} indicates that \textit{commit logs} contribute the most towards documentation, succeeding the \textit{source code comments} and \textit{textual documents}.
	        \newline
		     \newline
		     \fbox{\begin{minipage}{25em}
		    \textbf{RQ4:} Majority of the total documentation is extracted from source code comments (23.04\%), followed by textual documents (22.58\%). Commit logs and issues contributed to 18.5\% and 18.21\% of the total extracted documentation respectively. 17.63\% of the total extracted documentation was contributed by issues in the repositories. 
		    
		    \textbf{Insights-} Structured formats for each of these sources, could help in easy extraction and towards generation of unified documentation for a repository.
		    
		     \end{minipage}}

\section{Discussion}
Towards answering the research questions proposed, we realised that different documentation types could be derived from other software artifacts in the GitHub repositories. Researchers could explore the direction of consolidating this information across multiple artifacts to arrive at documentation for the repository. Establishing traceability links among documentation types and documentation sources is an important future direction. Tools towards generating documentation from multiple sources could be developed to ease the efforts of software practitioners. It might be difficult to generate meaningful documentation from the sources identified, requiring software practitioners to improve the textual information being logged through these artifacts. Also, there is a need for approaches and tools to migrate existing unstructured documentation into a structured form, for eliciting valuable information from multiple sources of software documentation. Furthermore, researchers could also explore specific formats for each of the identified documentation sources, to ensure extraction of information that could be easily converted to a meaningful documentation from the sources. 

\section{Threats to Validity}
\textbf{Internal Validity - }The categories in proposed taxonomy are based on developer interviews and card-sorting approach. Thus, there is a possibility of missing categories in the proposed taxonomy of types and sources of documentation, that might not have been observed during our interviews with 20 developers or during the execution of card-sorting approach. 

Moreover, as a first step towards considering documentation sources, we considered only pull requests, commits, issues, source code files and textual files which could be obtained through GitHub API or by downloading the repositories. Other sources such as Wiki, which are specific to GitHub projects, and, which require separate cloning, other than that of repository were not considered in this study. Such sources could be considered in the future versions of this study to improve the documentation being extracted. 

Except for the top 50 repositories in each programming language, topics of artifact data are labelled by comparing topic keywords with most frequently occurring keywords of respective labelled-topics, in top 50 repositories in each language. This generalization of keywords might not consider prominent keywords that could occur in other repositories, and thus could be biased towards the top 20\% repositories in each programming language. In addition, as we only analyzed textual sentences of data from \textit{issues}, \textit{pull requests} and \textit{commits}, we might have missed information from other data present in these artifacts such as \textit{date} and \textit{status}.
Also, only comments in source code files with extensions .cpp, .py, .cs and .java were analyzed, considering the complexity involved in addressing files of all programming languages. Also, text of files in the format specified in Table \ref{tab:fileclass} has been analyzed. As a result, useful insights on documentation that might be present in files of other formats such as \textit{.html, .tex, .pdf}, and comments in other programming language files are compromised.

\textbf{Construct Validity - }
All the results obtained are valid only for the dataset of considered 950 repositories. Performing this study on a different dataset might yield different results, However, considering the presence of repositories of varied programming languages in the dataset, indicating a representative sample of all repositories in the four programming languages, similar distribution of documentation types and sources could be identified in other repositories across GitHub, suggesting (but not proving) generalizability of the results.


\textbf{External Validity - }
The results obtained are confined to version of repositories and their and corresponding software artifacts as on 01 January 2021, as this empirical study is performed on locally downloaded repositories.

The accuracy of categorizing the extracted information into different documentation type classes is dependant on the efficiency of LDA approach. Different set of keywords might be identified for each class, consequently resulting in some sentences to be classified into different classes, if a different Natural Language Processing technique is employed.

\section{Related Work}
		\begin{table}[]
	    \caption{Documentation Sources and Types Observed from Literature}
	    \vspace{-2mm}
	    \centering
	    \begin{tabular}{|l|l|l|}
	    \hline
	         \textbf{Documentation Type} & \textbf{Source of Documentation} & \textbf{Literature}  \\
	         \hline
	         API-related  & example files, readme file & \cite{fowkes2016parameter} \cite{prana2019categorizing}
	         \\
	         \hline
	         Architecture-related & UML files, non-textual files & \cite{hebig2016quest}\\
	       
	         \hline
	         License-related  & commit messages & \cite{vendome2017license} \cite{prana2019categorizing}\\
	         & issues, License files, readme & \\

	         \hline
	         Error/Bug-related & issue comments & \cite{rastkar2010summarizing}\\
	         \hline
	         Project installation-related & readme& \cite{prana2019categorizing}\\
	         \hline
	         Project Enhancement-related & readme&\cite{prana2019categorizing}\cite{rastkar2010summarizing}\\
	         \hline
	         Project Background-related & readme&\cite{prana2019categorizing}\\
	         \hline
	         Project Status-related & readme, commits, issues,  &\cite{prana2019categorizing}\cite{coelho2018identifying}\\
	         &pull requests.&\\
	         \hline
	         Project Team-related & readme, commits, issues, &\cite{prana2019categorizing}\cite{rahman2014insight}\\
	         &pull requests.&\cite{coelho2018identifying}\\
	         \hline
	         Project Advantage-related & readme &\cite{prana2019categorizing}\\
	         \hline
	         References-related & readme&\cite{prana2019categorizing}\\
	         \hline
	         Contribution Guidelines & readme &\cite{prana2019categorizing}\\
	         \hline
	    \end{tabular}
	    
	    \label{tab:lit}
	\end{table}
	
	Repositories on GitHub comprises of various artifacts such as \textit{pull requests}, \textit{issue reports}, \textit{star-count}, \textit{fork-count}, \textit{number of watchers} and so on, that provide different types of information about the repository. Stars, forks and watchers are numeric in nature, indicating information on popularity of repositories \cite{borges2018s} whereas  software artifacts such as \textit{issue reports}, \textit{pull requests}, \textit{commit logs} contain information related to development and maintenance of the repositories \cite{dabbish2012social, sheoran2014understanding, rahman2014insight}.

	Researchers have conducted several studies to understand the role of these software artifacts as an aid for developers in improving quality and better handling of the project. A consolidated list of documentation types observed from the literature, along with their sources is presented in Table \ref{tab:lit}.

	\subsection{Using Individual Artifact Data}
	
	\textit{Issues}, \textit{commits} and pull \textit{requests} data is used in the literature for multiple research studies. Kikas et al. have observed various features present in \textit{issue reports}, as an aid in prediction of \textit{issue lifetimes} \cite{kikas2016using}. They have observed that analysis of contextual information from commits of the repository such as \textit{last commit date}, \textit{recent project activity}, \textit{date of issue creation} and other dynamic features such as number of actions performed on the issue, number of users who have worked on the issue, \textit{comment count} and so on, could help in predicting life time of the \textit{issue} \cite{kikas2016using}. 
	
	
	Liao et al. have observed that users who comment on \textit{issues} and who create \textit{pull requests} related to the \textit{issues} contribute more towards project development. 
	Zhang et al. have linked multiple related issues to support easy resolution and easy querying of issues by applying a combination of information retrieval technique and deep learning techniques \cite{zhang2020ilinker}. They also developed a tool to calculate frequency, word and document similarity scores between the queried issue and pending issues, based on the mentioned techniques to recommend related issues \cite{zhang2020ilinker}.
	

	Michaud et al. have attempted to identify branch of commits in the repository based on commit messages and types of merges to track the evolution history of a repository \cite{michaud2016recovering}. 
	
	Tsay et al. have presented a study that analyzed various factors of \textit{pull requests} that contribute towards their acceptance or rejection \cite{tsay2014influence}. It has been observed that pull requests with large number of changed files and more number of comments have lesser probability to be accepted, depending on other social factors \cite{tsay2014influence}. 

	\subsection{Using Integrated Data from Multiple Software Artifacts}
	Zhou et al. have considered various features from issue reports and commits to detect security threats or vulnerabilities in the repository \cite{zhou2017automated}. An optimal machine learning model has been built using commit messages from commits and multiple features of issue reports such as title, description, comments and so on, to detect unidentified security threats in the repository \cite{zhou2017automated}. 
	
	Considering the importance of information related to issue reports and commit history in assessing software quality and other factors of a repository, RCLinker has been proposed to link issue reports to their corresponding commits  by comparing source code of two consequent commits and summarizing the difference in source code \cite{le2015rclinker}. 
	
	Coelho et al. have considered features of multiple software artifacts such as issues, pull requests, commits and so on to identify status of maintenance of a project \cite{coelho2018identifying}. In an attempt to identify various issues in software documentation, Aghajani et al. have studied multiple sources that discuss about software documentation, which also include issue reports and pull requests on GiHhub, apart from developer email lists and discussions on knowledge sharing platforms \cite{aghajani2019software}. 
	
	\subsection{Documentation in Software Projects}
	A study has been performed by Borges et al. to identify various software artifacts that contribute to star count of GitHub repositories \cite{borges2018s}. It was observed that code quality and documentation largely influence star count. This study also re-emphasizes the importance of software documentation in GitHub repositories \cite{borges2018s}.
	
	Considering this importance of documentation, several attempts to improve and generate different types of software documentation are being developed. 
	
	\textit{Quasoledo} has been proposed to evaluate quality of documentation in \textit{Eclipse} projects, based on quality metrics related to completeness and readability \cite{schreck2007documentation}.
	Fowkes et al. have proposed an approach using probabilistic model of sequences to generate patterns of API, based on multiple usages in a project \cite{fowkes2016parameter}. 
Source code examples have been linked to official API documentation based on method calls and references to enable better API usage and understanding \cite{subramanian2014live}. A study has been performed by Hebig et al. to understand the number of projects in GitHub, that use uml diagrams \cite{hebig2016quest}. 


	In a survey with 146 software practitioners, Aghajani et al. have presented 13 documentation types, that were observed to be used by practitioners, for accomplishing multiple tasks \cite{aghajani2020software}. Of these 13 documentation types, around 6 of the documentation types correspond to end-user interactions with the project, such as \textit{video tutorials}, \textit{installation guides}, and so on. These set of documentation types also included \textit{Community Knowledge}, which could correspond to artifacts such as pull requests, issues and commits \cite{aghajani2020software} .

	
	
	Prana et al. have identified 8 categories of information present in ReadMe files through manual analysis of 50 ReadMe files, and annotated 150 ReadMe files based on these categories \cite{prana2019categorizing}.	
	Information in artifacts of the repository such as commit messages and discussions on issue trackers have been analyzed to obtain insights about changes in licenses used in the software \cite{vendome2015license}. Vendome et al. have attempted to trace the reasons for change in licenses through investigation of commit history and discussions in issue trackers and consequently obtaining insights on when (from commit history) and why (from issue tracker discussions) the licenses have been changed \cite{vendome2015license}.
	
	The existing literature has emphasized the need for software documentation, analyzed the availability of documentation types in GitHub repositories and has also presented approaches that consider information present in software artifacts, to generate respective documentation. In our analysis of the literature, we observe that the generation of documentation considers information from individual artifacts and is mostly limited to information present in \textit{bug reports}, \textit{source code} and \textit{readme files}. We also observe that documentation is present in many software artifacts, but is not explicitly mentioned. Extracting documentation information present in various software artifacts could help in enhancing the documentation. It is important to identify software artifacts that serve as potential sources of documentation. 
	However, to the best of our knowledge, we are not aware of any work in the literature that identifies and comprehends all the potential sources of different types of documentation. Hence, as a first step towards identifying potential documentation sources, we present a taxonomy of sources of documentation in GitHub. 
\section{Conclusion and Future Work}
In this paper, we proposed a taxonomy of documentation sources in GitHub repositories. We identified sources of documentation for different documentation types through card-sorting approach and developer interviews. This resulted in a taxonomy of six different documentation sources that could provide potential information for different documentation types identified. We also perform an empirical analysis to understand distribution of documentation types across multiple software artifacts that are identified as documentation sources, and the contribution of each source to the type of documentation. 950 GitHub repositories, with 233, 241, 253 and 223 repositories from C++, C\#, Java and Python programming languages respectively, have been scraped to perform the study. 

Software artifacts such as \textit{issues}, \textit{commits} and \textit{pull requests} of these repositories, not present explicitly in scraped data were fetched using GitHub API. Topic modelling has been applied on data extracted from all documentation sources and the obtained topics are labelled accordingly. The total extracted documentation from multiple sources, consisted 25.9\% of \textit{error-related} documentation and 23.6\% of \textit{project-related} documentation. \textit{File-related} documentation and \textit{license-related} documentation account to 16.04\% and 15.99\% of the total extracted documentation respectively. 5.63\% of the total extracted documentation consisted of \textit{API-related} documentation. \textit{Source code comments} contributed to 23.04\% and \textit{textual documents} contributed to 22.58\% of the total extracted documentation. \textit{Commit logs} and \textit{issues} contributed to 18.5\% and 18.21\% of the total extracted documentation respectively. 17.63\% of the total extracted documentation was contributed by \textit{issues} in the repositories. 


As a part of future work, we plan to identify other software artifacts that could be considered as potential sources of documentation. We also plan to improve the current taxonomy by adding more levels to the taxonomy. In future, information obtained from sources in the proposed taxonomy could be processed and used to generate documentation of different types. Also, techniques to extract text from design diagrams could be implemented, to leverage other documentation information that could be present in design diagrams.

	\balance
	
	\bibliographystyle{IEEEtran}
	\bibliography{IEEEabrv,samplbib}
	
\end{document}